\documentclass[aps,pra,preprint,onecolumn]{revtex4-2}
\usepackage{graphicx}
\usepackage{amsmath}
\usepackage{hyperref}
\hypersetup{hidelinks}
\makeatletter
\def\frontmatter@RRAP@format{\small\centering}
\makeatother

\newcounter{scwsubfigure}[figure]
\renewcommand{\thescwsubfigure}{\alph{scwsubfigure}}
\newenvironment{subfigure}[1]{%
  \refstepcounter{scwsubfigure}%
  \begin{minipage}{#1}\centering
  \renewcommand{\caption}[1]{\par\small(\thescwsubfigure) ##1\par}%
}{%
  \end{minipage}%
}

\begin{document}

\title{Performance of the subcarrier-wave quantum key distribution in
  the presence of spontaneous Raman scattering noise generated by
  classical DWDM channels}  

\author{F. Kiselev}
\email{kiselevfyodor@gmail.com}
\affiliation{ITMO University, Kronverkskiy, 49, St. Petersburg, 197101, Russia}

\author{R. Goncharov}
\affiliation{ITMO University, Kronverkskiy, 49, St. Petersburg, 197101, Russia}

\author{N. Veselkova}
\affiliation{ITMO University, Kronverkskiy, 49, St. Petersburg, 197101, Russia}

\author{E. Samsonov}
\affiliation{ITMO University, Kronverkskiy, 49, St. Petersburg, 197101, Russia}
\affiliation{Quanttelecom LLC., 6 Line, 59, St. Petersburg, 199178, Russia}

\author{A. D. Kiselev}
\affiliation{ITMO University, Kronverkskiy, 49, St. Petersburg, 197101, Russia}

\author{V. Egorov}
\affiliation{ITMO University, Kronverkskiy, 49, St. Petersburg, 197101, Russia}
\affiliation{Quanttelecom LLC., 6 Line, 59, St. Petersburg, 199178, Russia}

\begin{abstract}
In this paper we study performance of the subcarrier-wave quantum key
distribution system (SCW QKD) in the presence of spontaneous Raman
scattering (SpRS) noise  generated by classical channels of dense
wavelength division multiplexing (DWDM) network within a single-mode
optical fiber. We present the mathematical model for evaluation of the
quantum bit error rate (QBER) and the secure  key generation rate with
the SpRS noise taken into account. We consider two regimes of the SCW
QKD system: the continuous wave regime that uses continuous wave laser
and the pulsed regime. For these regimes, performance of the system is
analyzed depending on receiver sensitivity of classical DWDM. It is
found  that the pulsed regime outperforms the continuous wave regime
in both the secure key  generation rate and
the maximum achievable distance.  
\end{abstract}

\maketitle

%%%%%%%%%%%%%%%%%%%%%%%%%%  
\section{Introduction}
\label{sec:introduction}
%%%%%%%%%%%%%%%%%%%%%%%%%%%%%%%%%%%

Quantum key distribution (QKD) has become one of the most attractive
communication technologies in which the security is guaranteed by the
fundamental principles of quantum mechanics~\cite{Bennett1984,Gisin2002}. In the
last few decades, many attempts have been made to improve upon the
communication range and secure key generation rate of QKD.
These include the advanced single-photon detection
technology~\cite{He2017,Ma2016}, the remarkable QKD
protocols~\cite{Yin2018, Li2018}, and so on.
Considerable progress has been made to date:
the transmission distance of measurement-device-independent QKD in an
ultra-low-loss optical fiber can be as long as 404 km~\cite{Yin2016}
therewith quantum network can cover more than a 200-square-kilometer
metropolitan area ~\cite{Yin2017}. A high-speed QKD system with Mbit/s
secure key generation rate has been also achieved~\cite{Dynes2016}. 

The next step towards larger availability of QKD links is
to integrate QKD with existing fiber
infrastructures~\cite{Runser2007}. Common public dense wavelength
division multiplexing (DWDM) telecom networks can multiplex up to 50
different wavelength channels in a single fiber.
However,
such multiplexing
is a challenging task
as the quantum signal is much weaker than the classical
one. The launching power of 1~Gbps on-off keying (OOK) modulation
classical signal is typically 0~dBm, which is equivalent to
$8\times10^6$ photons per pulse~\cite{Niu2018},
whereas for the quantum
signal it is usually a few tenths of photons per pulse.
Due to the
large power gap between the two kinds of signals, the operation of QKD
would be severely degraded by noise impairments when using the same
fiber together with conventional data. So if the quantum channel is
launched into a fiber accompanied by other classical signals, several
effects, such as channel crosstalk, Raman scattering, four-wave mixing
(FWM) or amplified spontaneous emission (in the case of amplification
of the classical channels), can severely degrade the QKD system
performance.
A variety of methods suggested to suppress
the noise in DWDM-QKD systems
include
using narrow-band filters (NBFs)~\cite{Dynes2016,Peters2009,Eraerds2010, Xavier2011, Patel2014, Wang2015, Frohlich2017},
reducing the launching powers of classical
signals~\cite{Dynes2016,Eraerds2010, Xavier2011, Patel2014,Frohlich2017},
and the temporal filtering
technology~\cite{Dynes2016, Patel2014, Frohlich2017}. 

The first multiplexing scheme of QKD with classical signals was
implemented by Townsend in 1997~\cite{Townsend1997}.
In this scheme,
the classical signals are localized in the C-band and the quantum
signals are placed in the O-band through coarse wavelength-division
multiplexing (CWDM) components. The quantum signals  experience less
disturbances in this scheme as they are far away from the  classical
band at 1550~nm.
This CWDM-QKD scheme being widely accepted as a
viable solution has been the subject of intense studies~\cite{Chapuran2009,
  Aleksic2015, Choi2011, Wang2018, Mao2018}. 

However, the transmission loss in the O-band is much higher than that
in the C-band. DWDM components may thus be more advantageous.
It is also more compatible
with commercial optical network facilities,
so the DWDM-QKD
scheme has been attracted an increasingly large amount of attention in recent
years~\cite{Peters2009, Eraerds2010, Mora2012, Patel2014, Kumar2015,
  Bahrani2018}.

%%%%%%%%%%%%%%%%%%%%%%%%%%%%%%%%%%%%%%%%%%

% In this paper, we consider a quantum/classical channel
% WDM in the context of QKD in the presence of
% classical channels.
In this paper, we  carry out
a theoretical analysis and numerical simulation of the effect of
% a theoretical research and numerical
% simulation of the influence of
spontaneous Raman scattering (SpRS) on
the QKD channel in modern DWDM-QKD systems
caused by the presence of the forward and backward propagating classical signals
within a standard single-mode fiber.
In our analysis, all the channels are assumed to be located in the
narrow spectral C-band.

The impact of
various impairment sources (in particular, spontaneous Raman
scattering) in the DWDM-QKD 
systems was previously investigated
in~\cite{Kumar2015, Silva2014, Niu2018, Mlejnek2017,
  Eraerds2010, Sun2016, Bahrani2018}.
According to~\cite{Niu2018,
  Peters2009},
SpRS is the dominant source of noise for QKD in a DWDM
environment as long as the fiber length is beyond a few km.
The performance of the
DWDM-QKD systems with different channel allocation schemes
has been analyzed
for a variety of sifting protocols, such as BB84~\cite{Silva2014, Niu2018, Mlejnek2017, Eraerds2010},
SARG~\cite{Mlejnek2017, Eraerds2010},  COW~\cite{Mlejnek2017}, 
and Gaussian-modulated coherent state (GMCS) protocol~\cite{Kumar2015}.
The QKD systems utilizing the decoy-states~\cite{Silva2014},
and "plug and play" phase encoding~\cite{Eraerds2010}
were also considered in the context of the WDM.

The important parameters characterizing the quality of the QKD channel
are
the  quantum bit error rate (QBER) giving
the number of errors present in the key obtained after the sifting procedure
and the secure key generation rate which is
the rate at which the secret key is delivered to the recipient.
The expressions for these parameters
depend on the QKD protocol.
In the previous works~\cite{Kumar2015, Silva2014, Niu2018,
  Mlejnek2017, Eraerds2010}
QBER and the secure key rate have been studied
for various channel allocation schemes at
different system parameters such as the fiber attenuation, 
the fiber length, the launch power and the number of the classical channels,
the quantum receiver bandwidth, 
the classical channel modulation parameters.
Analytical results for both parameters
that take into account spontaneous Raman scattering
have been reported in~\cite{Kumar2015, Silva2014,  Mlejnek2017, Eraerds2010}.

% The important parameters characterizing the quality of the QKD channel
% are its QBER (quantum bit-error rate), the number of errors present in
% the key obtained after the sifting, and secure key generation rate,
% the rate at which the secret key is delivered to the recipient, whose
% expressions are defined by a particular QKD protocol. In the previous
% works on this subject~\cite{Kumar2015, Silva2014, Niu2018,
%   Mlejnek2017, Eraerds2010} QBER and secure key rate have been
% investigated under multiple channel allocation schemes and the
% different system parameters such as the fiber attenuation, fiber
% length, launch power and number of the classical channels, quantum
% receiver bandwidth, classical channel modulation parameters. Detailed
% expressions for both characteristics are given for various QKD
% protocols in the presence of spontaneous Raman scattering have been
% presented

We examine how the SpRS noise from classical DWDM
channels affects
the subcarrier-wave (SCW) quantum cryptography system~\cite{Gleim2017,Gaidash2019, Kiselev2020AnalysisSystem}
where,
in accordance with the quantum theory of electro-optic
phase modulation~\cite{Miroshnichenko2017},
each quantum channel can be
regarded as a pair of subcarrier waves (or a single
subcarrier~\cite{Kiselev2020AnalysisSystem}) resulting from the
phase modulation induced
transformation of monochromatic coherent light. 
Performance of a QKD
system based on SCW QKD system in the presence  of such noise
generated by classical channels of DWDM network operating within the same
optical fiber as the QKD system
has not been studied in any detail.

% This article is devoted to consideration of influence of SpRS noise
% from classical DWDM channels on a subcarrier-wave quantum cryptography
% system~\cite{Gleim2017, Gaidash2019}, where each quantum channel is
% regarded as a pair of subcarrier waves (or a single
% subcarrier) resulting from the
% transformation of monochromatic coherent light due to phase modulation
% in accordance with quantum theory of electrical-optical phase
% modulation.In this paper, performance of QKD
% system based on subcarrier-wave quantum key distribution system
% (SCW QKD) in the presence of such noise, generated by classical
% channels of DWDM network operating within the same optical fiber as
% QKD system,  was investigated  for the first time  ~\cite{Niu2018,
%   Mlejnek2017}. 

We consider two different regimes of
the laser source: the continuous wave (CW) regime and the pulsed regime.
We begin with a discussion of
the impact of the spontaneous Raman scattering
on the QBER
that will be computed using the approach developed in~\cite{Gaidash2019d}
for the given channel allocation scheme. 
Then we apply the Devetak-Winter approach to evaluate
the secure key generation rate
in the asymptotic limit of infinitely long keys.
By using
the results for QBER and the secure key generation rate,
we numerically calculate
the  maximum achievable distance
of the QKD system 
(the maximum distance at which the secret key rate is positive)
for different values of the receiver sensitivity of
classical channels.
The latter is defined as 
the minimum output power of the classical signal per channel
required to meet the BER requirements.
Finally, we
make a comparison between the two regimes of QKD system and
show that QBER, the secure key
generation rate and the maximum achievable distance are
significantly improved when
the continuous wave regime is changed to
the pulsed regime.

%  We also consider two different regimes of QKD system's laser:
%  continuous wave and pulsed. We discuss the impact of the spontaneous
%  Raman scattering on the DWDM-QKD system performance, namely, the
%  effect on the QBER (based on the approach developed
%  in) for the sub-carrier wave quantum key
%  distribution protocol for a given channel allocation scheme. 

% Then, based on the QBER and secure key generation rate, we numerically
% find the QKD system’s reach for different values of receiver
% sensitivity of classical channels, which defines the minimum output
% power of classical signal per channel required to satisfy BER
% requirements. Finally, we compare two regimes of QKD system and
% demonstrate how QBER, secure key generation rate and system's reach
% significantly improve once we move from continuous wave to pulse
% regime. 

%%%%%%%%%%%%%%%%%%%%%%%%%%%%%%%%
\section{Raman noise in fiber quantum channel}
\label{raman}
%%%%%%%%%%%%%%%%%%%%%%%%%%%%%%%%%%%

In this work we consider a DWDM system with the
channels that reside in
the C-band of the telecommunication window and
analyze its performance
depending on  a number of different parameters of
the DWDM system.
We shall use the model of SpRS presented in~\cite{ Mlejnek2017,
  Eraerds2010} where
the power of
the forward Raman scattering noise
induced by the presence of
classical channels can be written in the form:
\begin{equation}
\label{ram_pw_f}
P_{ram,f}=P_{out}L\sum_{c=1}^{N_{ch}}\rho (\lambda _{c},\lambda _{q})\Delta \lambda. 
\end{equation}

This is the case
where the signals in quantum and classical channels
are propagating in optical fiber
along the same direction. 
In the opposite case of
backward Raman scattering noise
where
the signals are counter-propagating
the power is given by
\begin{equation}
\label{ram_pw_b}
    P_{ram,b}=P_{out}\frac{\sinh(\xi L)}{\xi}\sum_{c=1}^{N_{ch}}\rho (\lambda _{c},\lambda _{q})\Delta \lambda, 
\end{equation}
where $P_{out}$ is the output from the fiber
power of a single classical channel,
$\xi$ is the attenuation of the
fiber,
$L$ is the length of the optical fiber,
$N_{ch}$ is the number of classical channels present in DWDM system,
$\rho (\lambda_{c},\lambda _{q})$
is the normalized scattering cross-section for the
wavelengths of classical ($\lambda _{c}$) and quantum ($\lambda _{q}$)
channels
(we shall use the data presented in Figure~1 of~\cite{Eraerds2010}),
and $\Delta \lambda$ is the bandwidth of the quantum channel filtering
system.

The output power is chosen so as to meet the BER conditions
for classical communications~\cite{Mlejnek2017}.
It is determined by the
receiver sensitivity $R_{x}$ and the insertion losses $IL$ of the
system
as follows
\begin{equation}
    P_{out}[\mathrm{dBm}]=R_{x}[\mathrm{dBm}]+IL[\mathrm{dB}].
\end{equation}

In this paper we consider the classical receiver sensitivity,
which is the minimum power required at the receiver of the classical
channel to detect the signal reliably,
varying from
$-23$~dBm, which is typical for OOK modulation formats and thus
1~Gbps networks,
to $-48$~dBm, which corresponds
to a more modern 40G or 100G networks with
binary phase-shift keying
(BPSK) formats and coherent detection~\cite{Mlejnek2017,kikuchi2008evaluation}.
In addition, according to~\cite{Mlejnek2017},
for higher data rate optical systems,
the receiver sensitivity
can be even higher.

Given the power of Raman noise in the quantum channel,
we can calculate the probability of noise induced photon detection for
both forward and backward Raman scattering.
By assuming that
SpRS noise is evenly distributed in time,
the total energy of SpRS light within the detector gating time $\Delta
t$
can be computed
as a product of
$\Delta t$
and
the power calculated either in
Eq.~\eqref{ram_pw_f} or in Eq.~\eqref{ram_pw_b}. 
Then the result divided
by the energy of a single photon
at the wavelength of quantum channel
gives the mean photon number.
The detection probability of SpRS photon
can now be calculated by multiplying the mean photon number
by the detector
efficiency $\eta_D$ and
the transmission coefficient associated with the
insertion losses $\eta_{B}=10^{-0.1IL}$ of the detection
system.
The final result reads 
\begin{equation}
\label{ram_prob}
    p_{ram,f/b}=\frac{P_{ram,f/b}}{hc/\lambda_{q}}\Delta t\eta_{D}\eta_{B},
\end{equation}
where $h$ is the Planck constant
and $c$ is the speed of light.
This probability
can now be used to estimate the impact of the SpRS
noise on the performance of the QKD system.  
Parameters of the DWDM system are summarized in Table \ref{tab1}. 

\begin{table}[ht]
\centering
\begin{tabular}{|c|c|}
\hline
Parameter                                    & Value                     \\ \hline
$\xi$                           & 0.18 dB/km    \\ \hline
$\Delta \lambda$ & 15 GHz                    \\ \hline
$N_{ch}$                                    & 40                        \\ \hline
$R_{x}$                                     & from $-23$ dBm to $-48$ dBm     \\ \hline
IL                                           & 8 dB                       \\ \hline
$\lambda_{q}$               & 1535 nm                    \\ \hline
$\lambda_{c}$               & from 1548.5 nm to 1564.3 nm \\ \hline
\end{tabular}
\caption{Parameters of DWDM system and allocation of quantum channel}
\label{tab1}
\end{table}

%%%%%%%%%%%%%%%%%%%%%%%%%%%%%%
\section{SCW QKD protocol and setup}
\label{sec:SCW QKD-protocol}
%%%%%%%%%%%%%%%%%%%%%%%%%%

\begin{figure}[ht]
   \centering
     \includegraphics[width=\textwidth]{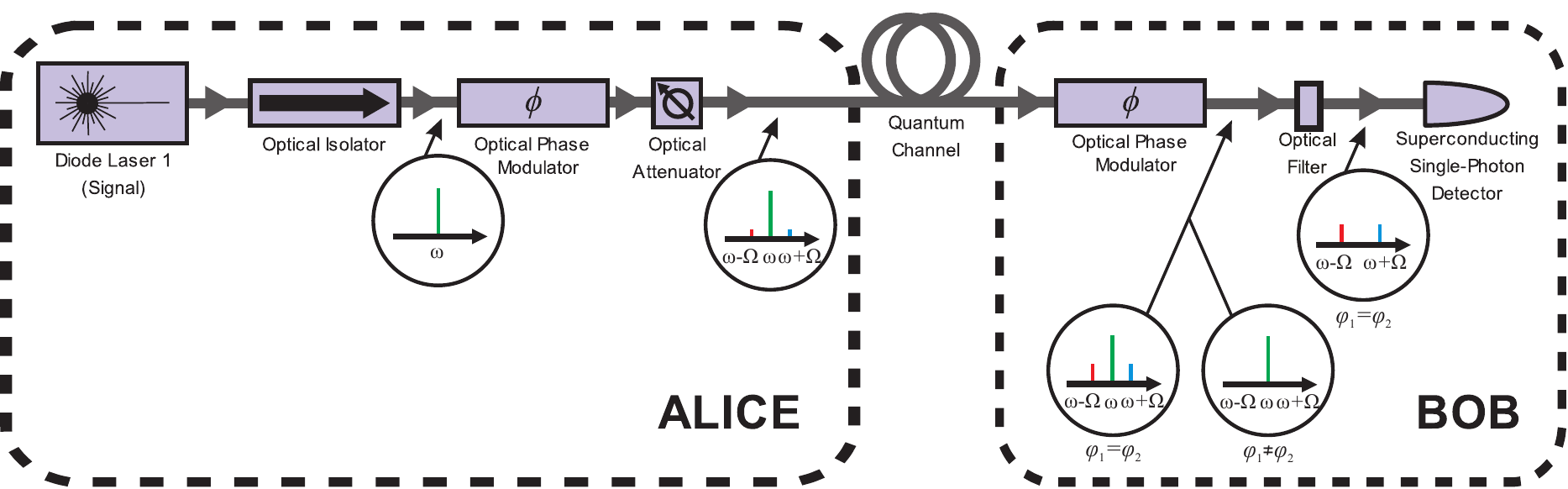}
\caption{Schematic of the subcarrier-wave quantum key distribution
  system. Insets (in circles) show the simplified intensity
  spectra. An optical isolator is required to prevent reflection of
  the beam light, an attenuator is required to achieve the needed mean
  photon number in the quantum channel.} 
\label{scheme}
\end{figure}

 In the SCW version of BB84
 protocol~\cite{Gleim2016,miroshnichenko2018security} (the number of
 bases is
 $N=2$) in Alice's block a coherent monochromatic light beam at optical frequency
 $\omega$ is modulated in a single-tone traveling-wave phase modulator
 with the modulation frequency $\Omega$ and the phase $\varphi_{A} \in \{\{0,\;\pi\},\;\{\pi/2,\;3\pi/2\}\}$.  
During the modulation process the energy transfers from central mode
to $2S$ vacuum sideband modes forming a resultant signal at
frequencies $\omega_j=\omega+j\Omega$ ($-S \leq j \leq S$).
Signal amplitudes can be expressed in terms of Wigner
d-functions~$d_{n j}^{S}(\beta)$~\cite{Miroshnichenko2017,
  varshalovich1988quantum}. In the transmission window $T$ (which in CW case is the duration of the signal per bit, and in pulsed case the duration of the pulse) the average number of photons $\mu_0$ and the modulation index~$m$ are chosen
so as to maximize the secure key generation rate, and the
phase~$\varphi_{A}$ encodes a random bit value.  The signal passes the quantum channel suffering losses
$\eta(L)=10^{\frac{-\xi L}{10}}$, and enters the Bob's module, where
it is modulated again with the same modulator setup, except a
different random phase shift~$\varphi_{B}$, so the final signal
depends on the phase difference $\Delta
\varphi=\varphi_{A}-\varphi_{B}$.
Further, only the sidebands are collected by
the detector. The carrier mode is pre-cut with a spectral
filter with the attenuation coefficient~$\vartheta$. The experimental setup of SCW
QKD system is shown in Figure~\ref{scheme}. A detailed description of the protocol can be
found in~\cite{miroshnichenko2018security,kozubov2019finite}.
Here we briefly describe equations
needed for subsequent calculations.  

The mean number of photons arriving on the single-photon detector over
the transmission window~$T$ can be written
in the following form~\cite{miroshnichenko2018security}: 
\begin{align}
  \label{eq:n_ph}
    n_{p h}\left(\varphi_{A}, \varphi_{B}\right)
  =\mu_{0}
  \eta(L)
  \eta_{B}\left(1-(1-\vartheta)\left|d_{00}^{S}\left(\beta^{\prime}\right)\right|^{2}\right)
\end{align} 
with the relations  
\begin{align}
  &
    \cos \beta^{\prime}\equiv
1-\frac{1}{2}\left(\frac{m'}{S+0.5}\right)^{2}
    =\cos ^{2} \beta-\sin ^{2} \beta \cos
  \left(\varphi_{A}-\varphi_{B}\right),\:
    \cos\beta=1-\frac{1}{2}\left(\frac{m}{S+0.5}\right)^{2}
\end{align}
that determine the angle $\beta^{\prime}$.
Note that, in  the limit of large $S$ and small modulation index,
we can use the approximate expression for
$d_{00}^{S}(\beta^{\prime})$:
\begin{align}
  \label{eq:d_approx}
  d_{00}^{S}(\beta^{\prime})
  \approx
  J_0(m')\approx 1-(m')^2/4,
\end{align}
where 
$(m')^2=2m^2(1+\cos(\varphi_A-\varphi_B))$
and $J_0(m')$ is the zero-order Bessel function of the first kind.

%%%%%%%%%%%%%%%%%%%%%%%%%%%%%%%%%
\section{Raman noise in SCW QKD}
\label{sec:raman-noise}
%%%%%%%%%%%%%%%%%%%%%%%%%%%%%%%%%% 

%%%%%%%%%%%%%%%%%%%%%%%%%%%%%%%%%%%%%%%%%%%
\subsection{Quantum bit error rate}
\label{subsec:qber}
%%%%%%%%%%%%%%%%%%%%%%%%%%%%%%%%%%%%%%%%%%%%

Since the average photon number in sidebands
is small (much less than unity),
one can use the theory of L.~Mandel~\cite{Mandel1995} which
describes the probability of a single photon detector click in the linear approximation.
The contribution of Raman scattering in fiber channel is described by
the average photon fractions that can be taken into account as an additional
term in the expression for the probability of the detector to produce a click during the gating time $\Delta t$ 
\begin{equation}
P_{d e t}\left(\varphi_{A}, \varphi_{B}\right)=\left(\eta_{D}
  \frac{n_{p h}\left(\varphi_{A}, \varphi_{B}\right)}{T}+
  \gamma_{d a r k}\right) \Delta t + p_{ram}
=
p_{cl}(\varphi_A,\varphi_B)+
p_{d a r k}+p_{ram},
\label{eq:Pdet}
\end{equation}
where $\gamma_{d a r k}$ is the dark count rate,
$p_{d a r k} \equiv \gamma_{d a r k} \Delta t$,
and the additional probability term $p_{ram}$
depends on the type of scattering as follows   
\begin{align}
&
                \text{forward: } p_{ram}=p_{ram,f},
                \label{eq:forward}
  \\
  &
    \text{forward+backward (full): } p_{ram}=p_{ram,f}+p_{ram,b}
\label{eq:forw-backward}
    .
\end{align}  

Let Alice choose $\varphi_{A} = 0$.
Assuming that
Bob guesses the correct basis
(results with wrong bases will be discarded at the sifting stage),
the probability of error
is the probability to obtain a click
at $\varphi_B = \pi$.
Then,
following~\cite{miroshnichenko2018security}, 
we can define the parameters of the binary symmetric error and erasure (BSEE)
channel~\cite{Cover2006}
between Alice and Bob as follows
\begin{align}
  &
    E=P_{det}(0, \pi+\delta \varphi),
    \label{eq:E}
    \\
    &
      1-G-E=P_{d e t}(0,\delta \varphi),
      \label{eq:G}
  \end{align}
      where
      $G$ is the conditional probability of receiving an inconclusive
      measurement result;
      $E$ is the conditional probability of an incorrect
      bit measurement;
      and $\delta\varphi$ is the apparatus related phase mismatch.
      These parameters determine QBER
\begin{equation}
Q=\frac{E}{1-G}=\frac{P_{d e t}(0, \pi+\delta \varphi)}{P_{d e t}(0,
  \delta \varphi)+P_{d e t}(0, \pi+\delta \varphi)}
\label{eq:QBER}
\end{equation}
that gives the probability for Bob to accept an erroneous bit.

      In our calculations,
      we shall use
      the approximation~\eqref{eq:d_approx}
      giving
      an approximate version of the above expression
\begin{equation}
Q=\frac{2\mu \tau \eta(1-\vartheta)\left(1-\cos (\delta
    \varphi)\right)+\tau \vartheta \mu_{0} \eta+p_{d a r
    k}+p_{ram}}{4 \mu \tau
  \eta(1-\vartheta)+2\tau \vartheta \mu_{0}
  \eta+2 p_{d a r k}+2p_{ram}},
\label{eq:QBER-approx}
\end{equation}
where $\eta \equiv \eta_{B} \eta(L) \eta_{D}$,
$\mu=\mu_0 m^2$
and
$\tau \equiv\Delta t/T$.

In what follows, we consider
two possible
regimes of radiation source: the pulsed regime and
the continuous wave regime.
The latter corresponds to the experimental
model implemented so far in which the continuous radiation is
modulated at the given repetition rate.
In this regime, the coherent state that corresponds to one bit of
information, similar to the SpRS noise,
is evenly distributed in time along the phase modulation period. 
As a result, in this case,
the signal to noise ratio is solely governed by 
the average photon number of this state.
For small gating time, it additionally results in
reduction of the ratio $\tau=\Delta t/T<1$.
%that has a detrimental effect on the performance.
%And the QBER is independent on the gating time.
By contrast, in the pulsed regime,
the pulse width can be changed so as
to have the transmission window equal to the gating time with $\tau=1$.
As a consequence, the signal to noise ratio will increase.
Note that, in both regimes, the repetition rate of random phase shifts is the
same. 

In our calculations, we have used the following parameters:
$\Omega=4.8$~GHz,
$\mu_0=3.93$,
$m=0.319$,
$\xi=0.18$~dB/km,
$\vartheta=10^{-3}$,
$\Delta t = 1$~ns,
$\eta_{D}=0.1$,
$p_{dark}=4\times10^{-6}$,
$\delta\phi=5^{\circ}$. Bob's module losses are 8~dB.
For the pulsed and continuous regimes,
$T =1$~ns and $T = 10$~ns, respectively. 

\begin{figure}[ht]
\centering
\begin{subfigure}{.5\textwidth}
  \centering
  \includegraphics[trim=20 0 20 10,width=\linewidth]{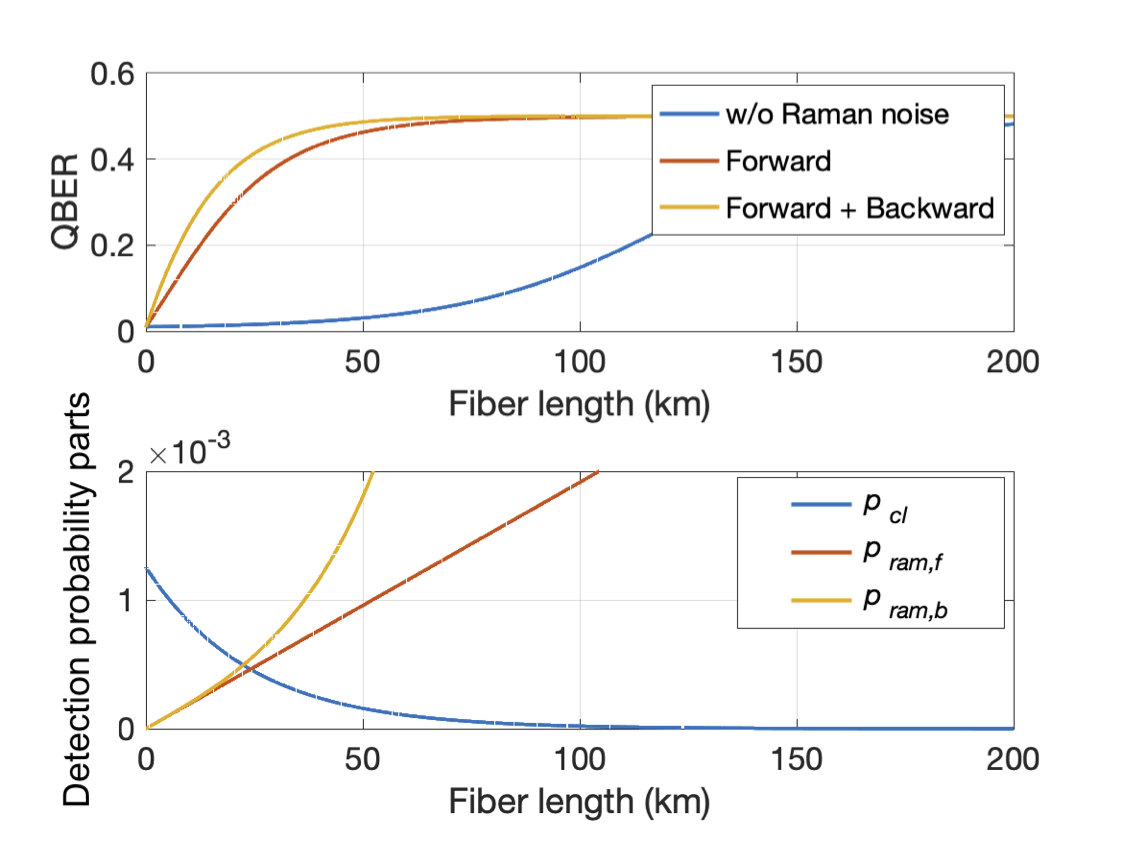}
  \caption{}
  \label{qber-cw}
\end{subfigure}%
\begin{subfigure}{.5\textwidth}
  \centering
  \includegraphics[trim=20 0 20 10,width=\linewidth]{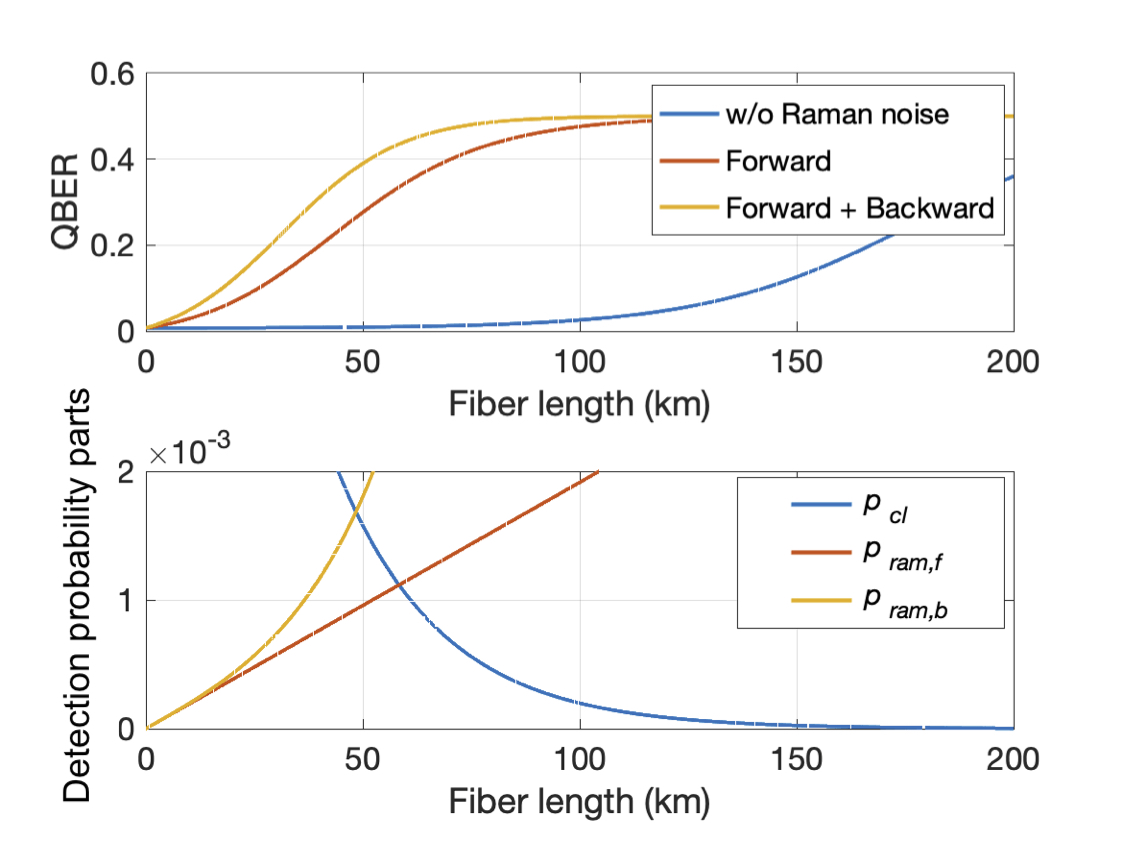}
  \caption{}
  \label{qber-p}
\end{subfigure}
\caption{QBER and contributions to the probability of detection $P_{det}$ (see Eq.~\eqref{eq:Pdet}) versus optical fiber
  length for continuous wave (a) and pulsed (b) laser regimes. The
  sensitivity of the classical channel receiver is $-28$~dBm.} 
\label{qber}
\end{figure}

Figure~\ref{qber} shows
dependence of QBER on
the fiber length computed
for the three cases:
the case without SpRS noise with $p_{ram}=0$,
and the two cases of Raman scattering with $p_{ram}$
described in Eqs.~\eqref{eq:forward} and~\eqref{eq:forw-backward}.
For comparison, we demonstrate the
results  of operation in (a)~the continuous wave and
(b)~the pulsed regimes. 

It can be seen that,
in the presence of SpRS noise,
QBER rapidly grows
quickly reaching the critical value ($\approx 0.07$)
at which the secret key cannot be generated (see
Section~\ref{subsec:comp-eval-secr}).
Clearly, the fastest growth
occurs in the case of full scattering,
% However, with a fairly low channel's
% sensitivity, the visual difference is not that noticeable.
An important point is
that there is a considerable difference between
the continuous
wave and the pulsed modes of operation.
In particular, in the CW regime,
the increase in the number of errors is so fast that
the convex parts of the curves
disappear.
There also is a noticeable difference
between the curves
for these modes
representing the results
without the SpRS noise.

Referring to Fig.~\ref{qber},
the  click probability
of detection,
$p_{cl}\equiv p_{cl}(\varphi_A,\varphi_A)$,
(see Eq.~\eqref{eq:Pdet})
for the pulsed regime
is ten times larger than
this probability
in the continuous regime.
Obviously,
the reason why
the pulsed regime shows much
better results is that the signal entirely fits into the gate.

% \begin{figure}[ht]
% \centering
% \includegraphics[width=\linewidth]{28dbm_sens_Q_cw_IL.eps}
% \caption{QBER and photon count probability versus quantum channel length for continuous wave laser regime and classical channel receiver sensitivity of 28dBm.}
% \label{qber}
% \end{figure}

% \begin{figure}[ht]
% \centering
% \includegraphics[width=\linewidth]{28dbm_sens_Q_IL.jpg}
% \caption{QBER and photon count probability versus quantum channel length for pulsed wave laser regime and classical channel receiver sensitivity of 28dBm.}
% \label{qber}
% \end{figure}

%%%%%%%%%%%%%%%%%%%%%%%%%%%%%%%%%%%%%
\subsection{Secret key generation rate} 
\label{subsec:comp-eval-secr}
%%%%%%%%%%%%%%%%%%%%%%%%%%%%%%%%%%%%%%

In assessing the security under the influence of SpRS noise,
we restrict our analysis to the asymptotic case
of the keys of infinite length.
Then,
for one-way QKD protocols
with independent identically distributed information carriers
the secure generation rate
in the presence of collective attacks
is lower bounded by the Devetak-Winter
bound~\cite{devetak2005distillation}
\begin{equation}
    \label{Keq}
    K=v_{S} P_{B}\left[1-\operatorname{leak}_{E C}(Q)-\max _{E} \chi(A: E)\right],
\end{equation}
where $v_{S}$ is the repetition rate (in our system $v_{S}=100$~MHz),
$P_B=(1-G)/N$ is the probability of successful state detection in the
guessed basis ($N$ is the number of bases), the amount of information disclosed by Alice during
error correction, $\operatorname{leak}_{E C}(Q)$, is
limited by the Shannon bound:
$\operatorname{leak}_{E C}(Q)\geq h(Q)$, where $h(x)$ is the binary
entropy; and the last term is the Holevo information.

We shall assume that Eve is not affected by Raman scattering
and calculate the Holevo bound using the results of
Ref.~\cite{miroshnichenko2018security}
derived for collective beam-splitting attacks.
According to Ref.~\cite{miroshnichenko2018security},
the secure key generation rate~\eqref{Keq}
can be estimated as follows
\begin{equation}
  \label{eq:K_rate}
K=\frac{(1-G)v_{S}}{2}\left[1-h\left(Q\right)-h\left(\frac{1-e^{-\mu_{0}
        m^2}}{2}\right)\right],
\end{equation}
where we have used the approximation~Eq.~\eqref{eq:d_approx}
to simplify the expression for
the $d$ function: $d_{00}^{S}(2\beta)\approx 1-m^2$.

\begin{figure}[ht]
\centering
\begin{subfigure}{.5\textwidth}
  \centering
  \includegraphics[width=\linewidth]{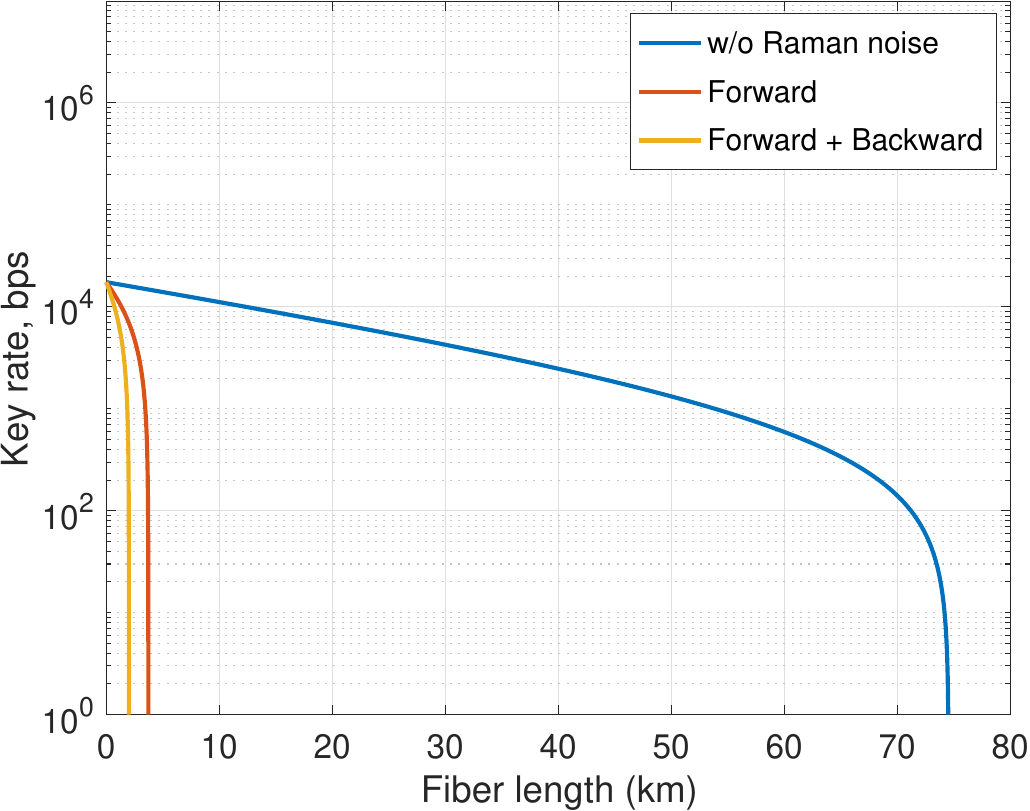}
  \caption{}
  \label{Kfull-cw}
\end{subfigure}%
\begin{subfigure}{.5\textwidth}
  \centering
  \includegraphics[width=\linewidth]{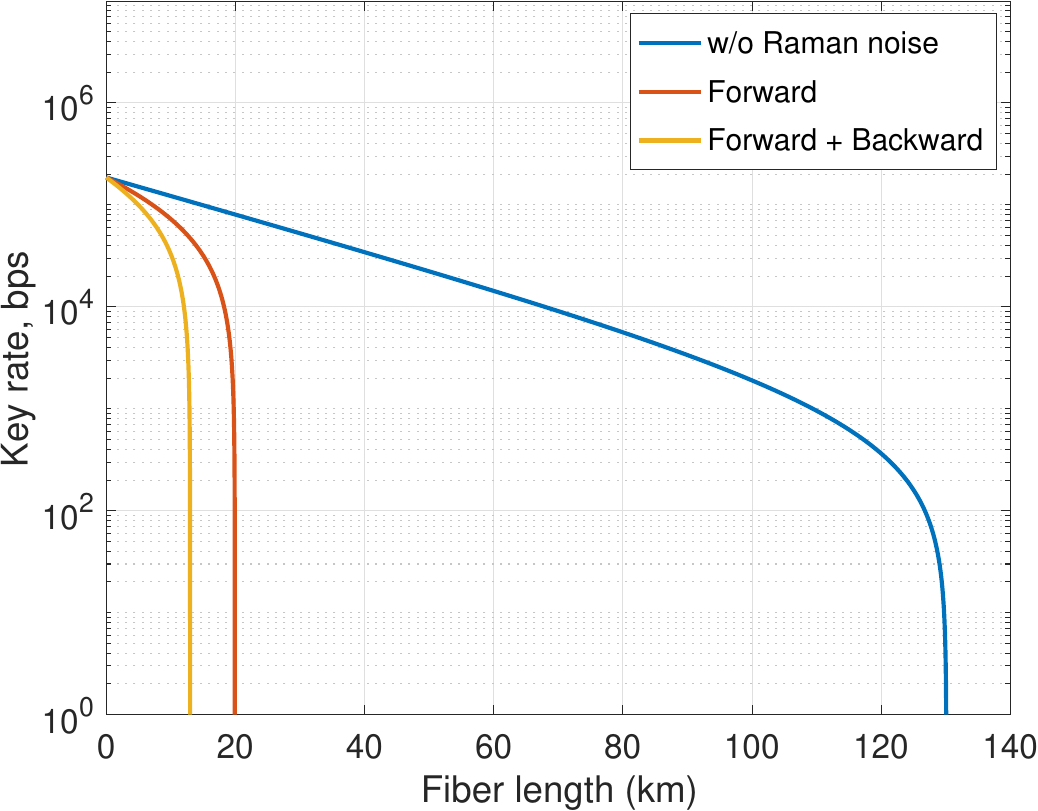}
  \caption{}
  \label{Kfull-p}
\end{subfigure}
\caption{Secure key generation rate versus optical fiber length for
  continuous~(a) and pulsed~(b) laser regimes. The sensitivity of
  the classical channel receiver is $-28$~dBm.} 
\label{Kfull}
\end{figure}

The SpRS noise is found to have a profound effect on QBER,
so it will also affect the secret key rate~\eqref{eq:K_rate}.
Figure~\ref{Kfull} presents the results for
the key rate computed as a function of the fiber length.
As compared to the case without the SpRS noise, 
the curves with the SpRS noise taken into account
indicate a pronounced drop of the maximum
achievable distance.
This distance decreases dramatically at the forward scattering
and is further reduced at the full scattering.
Importantly,
in the pulsed regime, the rate and the distance are both
much higher than those in the continuous regime.  

% \begin{figure}[ht]
% \centering
% \begin{subfigure}{.5\textwidth}
%   \centering
%   \includegraphics[width=\linewidth]{QBER+K-cw.eps}
%   \caption{}
%   \label{QBER+K-cw}
% \end{subfigure}%
% \begin{subfigure}{.5\textwidth}
%   \centering
%   \includegraphics[width=\linewidth]{QBER+K-p.eps}
%   \caption{}
%   \label{QBER+K-p}
% \end{subfigure}
% \caption{Secure key generation rate and QBER versus quantum channel length for continuous~(a) and pulsed~(b) laser regimes. Distillation limit is set to 0.05. Classical channel input power is -19~dBm.}
% \label{double}
% \end{figure}

\begin{figure}[ht]
\centering
\begin{subfigure}{.5\textwidth}
  \centering
  \includegraphics[width=\linewidth]{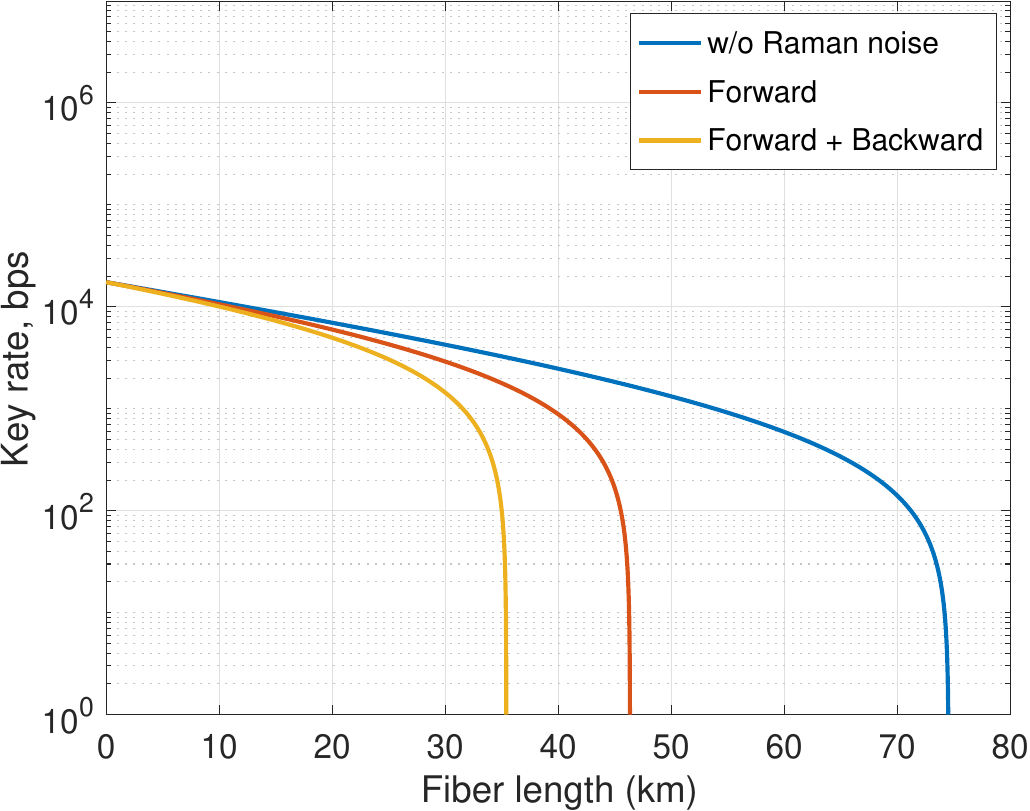}
  \caption{}
  \label{Kfull2-cw}
\end{subfigure}%
\begin{subfigure}{.5\textwidth}
  \centering
  \includegraphics[width=\linewidth]{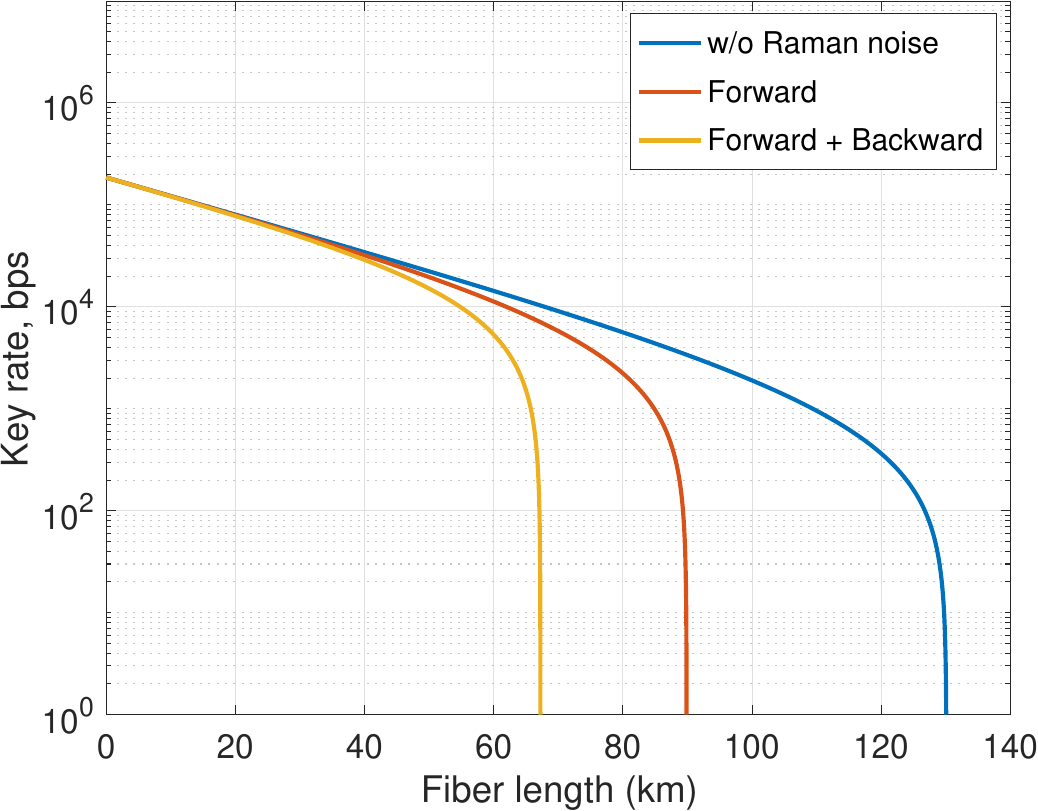}
  \caption{}
  \label{Kfull2-p}
\end{subfigure}
\caption{Secure key generation rate versus optical fiber length for
  continuous~(a) and pulsed~(b) laser regimes. The sensitivity of
  the classical channel receiver is $-48$~dBm.} 
\label{Kfull2}
\end{figure}

Figure~\ref{Kfull2} shows
what happen to the fiber length dependence of
the key generation rates
when
the receiver sensitivity
of the classical channel
is lowered down to $-48$~dBm.
It can be seen that
such change of the sensitivity value
leads to an increase in the maximum distance.
In this case,
effects the Raman
scattering on the SCW system 
are less pronounced.
It is noteworthy to note that the pulsed
regime still provides the best performance.  

\begin{figure}[h!
t]
\centering
\includegraphics[width=250pt]{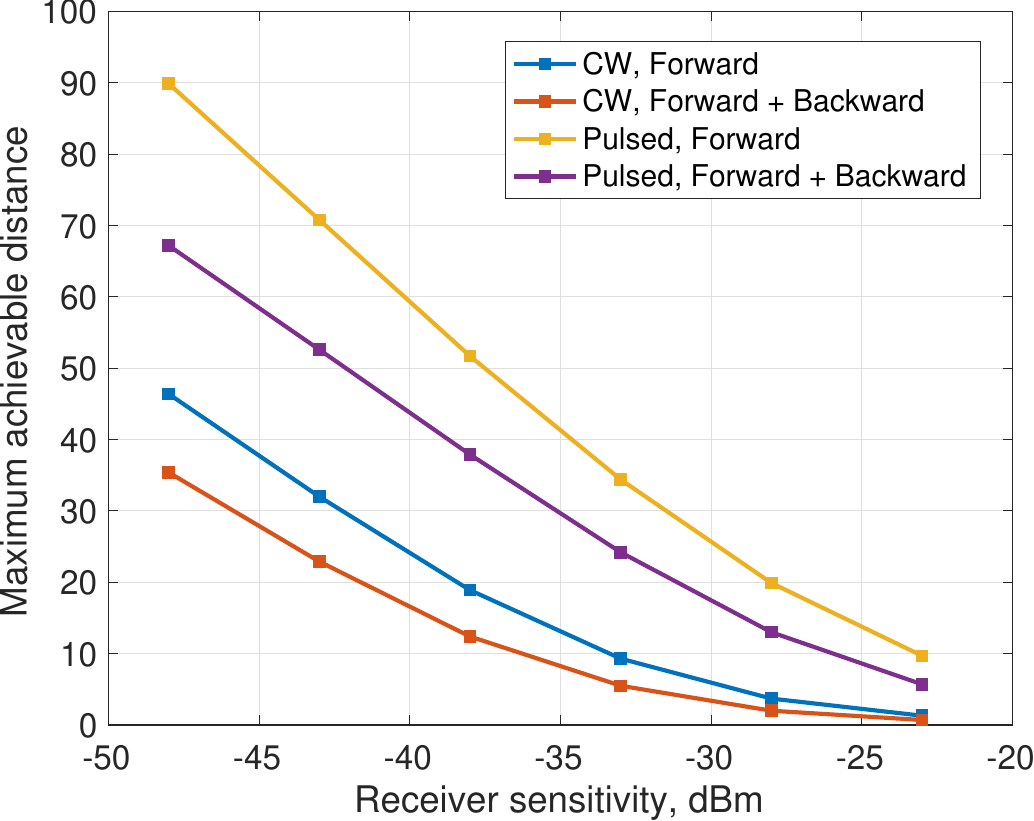}
\caption{Maximum possible distance at which the SCW QKD system can
  operate versus classical channel receiver sensitivity.} 
\label{reach}
\end{figure}

Effects of the receiver sensitivity on
the achievable maximum distance
are summarized in Figure~\ref{reach}.
The distance is shown to be
a decreasing function
of the sensitivity
and the best results
are obtained in the case of the pulsed regime.

%%%%%%%%%%%%%%%%%%%%%
\section{Conclusion}
\label{sec:conclusion}
%%%%%%%%%%%%%%%%%%%%%%

In this paper we have studied how the SpRS noise generated by the
classical traffic affects the performance of the SCW QKD system.
We have considered two different types of SCW QKD setups
that operate using either the CW laser radiation or
the pulsed regime of radiation.
Both types
are modulated at the same frequency of 100~MHz.
In the presence
of SpRS noise,
the pulsed version of the setup
is shown to
perform much better
than the CW laser radiation
both in terms of
the maximum achievable distance and
the secure key generation
rate.
In particular, we have found that,
for the receiver sensitivity of
$-28$~dBm,
the maximum distance
is about
4~km
when
the CW version of the setup is used.
This distance is a typical high-end value for the 1~Gbps network with
OOK modulation.
At the same time,
for the same parameters of
classical DWDM network,
using the pulsed regime
allows to achieve the maximum distance of about 20~km.
We have also shown that
further improvements
can be made
by
reducing
the receiver
sensitivity down to $-48$~dBm.
In this case,
the maximum distance
for the pulsed SCW QKD
increases up to 90~km in the presence
of SpRS noise from classical DWDM traffic.
This result is typical for 40G-100G networks  with
coherent detection.        

\section*{Acknowledgements}
The work was done by Leading Research Center "National Center of
Quantum Internet" of ITMO University during the implementation of the
government support program, with the financial support of Ministry of
Digital Development, Communications and Mass Media of the Russian
Federation and RVC JSC; Grant Agreement ID: 0000000007119P190002,
agreement No. 006-20 dated 27.03.2020

\bibliography{references}
\end{document}